\documentstyle[aps,manuscript]{revtex}

\makeatletter

\makeatletter
\makeatletter
\makeatother
\makeatother

\makeatother

\begin{document}

\title{Relativistic theories of interacting fields and fluids }

\author{Esteban Calzetta\thanks{
email: calzetta@df.uba.ar
} and Marc Thibeault\thanks{
email: marc@df.uba.ar
}\\
{\small Departamento de Física, UBA, Buenos Aires, Argentina}\small }

\maketitle
\begin{abstract}
We investigate divergence-type theories describing the disipative interaction
betwen a field and a fluid. We look for theories which, under equilibrium conditions,
reduce to the theory of a Klein-Gordon scalar field and a perfect fluid. We
show that the requirements of causality and positivity of entropy production
put non-trivial constraints to the structure of the interaction terms. These
theories provide a basis for the phenomonological study of the reheating period. 
\end{abstract}

\section{Introduction}

In this paper we investigate divergence-type theories (DTT) describing the dissipative
interaction between a field and a fluid \cite{Geroch,Calzetta}. We look for
theories which, under equilibrium conditions, reduce to the theory of a Klein-Gordon
(KG)scalar field and a perfect fluid. We show that the requirements of causality
and positive entropy production put non-trivial constraints on the structure
of the interaction terms. 

The motivation for this work comes from the development of inflationary cosmological
models \cite{Guth,LindeA,KolbTurner}. In these models of the Early Universe,
most of the energy density is concentrated in a single (fundamental or effective)
scalar field or ''inflaton''. There are two well defined moments in the evolution,
the roll - down period and the reheating period. In the former the inflaton
provides an effective cosmological constant and supports the superluminal expansion
of the Universe \cite{TurnerHu}; in the later the inflaton decays into ordinary
matter, thus creating entropy and heating up the Universe. 

While inflationary models assume a simple spatial dependence for both the inflaton
and the metric of the Universe, scalar and gravitational fluctuations around
this simple background play an important role \cite{BrandenbergerMukhanov}.
During roll - down, scalar and tensor fluctuations are redshifted and frozen
by the cosmological expansion, and become the seeds for future contrasts in
the energy density of the Universe. During reheating, fluctuations within the
horizon interact in a complex way with the background fields, and are key to
such questions as what is the duration of the reheating period, at which temperature
is equilibrium attained, how is energy distributed at the end of reheating,
and the spectrum of primordial density fluctuations. 

Overall, quantum field theory in curved spaces is an adequate framework to study
the roll - down period (although of course many questions remain, such as what
is the shape of the inflaton potential, and whether the fine tuning usually
assumed in standard inflationary models is truly unavoidable)\cite{CKLL,MTurner,reconstruction}.
However, reheating is much more complex, as here neither background fields nor
fluctuations may be considered a perturbation on the other, and moreover the
way the geometry changes along the process has a drastic influence on results.
For this reason, analysis based on quantum field theory on curved spaces have
been satisfactory only for the early stages of reheating, the so-called preheating
where the dominant process is particle creation from the background field through
parametric amplification \cite{Linde94,Linde97,Khlebnikov1,K2,Boyanovsky00,Brandenberger00,RamseyHu97a,RamseyHu97}.

To the best of our knowledge, the most successful strategy to deal with reheating
in the strongly nonlinear regime has been to describe the inflaton field(s)
as purely classical \cite{Felder}. By modelling the inflaton as a classical
field obeying a non linear wave equation it is possible to go beyond perturbation
theory; moreover, this approach may be justified on the basis that occupation
numbers, in the infrared modes of interest, are typically very high. However,
a purely classical theory ignores virtual processes that may modify the behavior
of the quantum theory, as has been shown by recent detailed calculations of
transport coefficients from quantum field theory \cite{Kadanoff,J1,J2,JeonYaffe,Carrington99,CalzettaHu}. 

The next step in improving classical models of reheating is to incorporate those
quantum aspects as phenomenological terms within the classical theory \cite{Hu}.
For example, the inflaton and its fluctuations may be depicted as a classical
dissipative fluid, with constitutive relationships derived in some way from
quantum field theory. 

The simplest and first approach to reheating, for example, is to transform the
Klein - Gordon equation into a telegraphist's equation by adding a \( \Gamma \dot{\phi } \)
dissipative term, where \( \Gamma  \) is estimated from the quantum mechanical
decay rate of the inflaton (\cite{Abbott,Albrecht,Dolgov1,Dolgov2}). As pointed
out by Brandenberger and others\cite{Brandenberger99}, this approach misses
some important components of the reheating process, such as the possibility
of preheating, namely, enhanced decay through parametric amplification of quantum
fields coupled to the inflaton. A subtler criticism is that this approach, or
rather, simple covariant generalizations of this approach, are equivalent to
writing a first order dissipative theory for the inflaton field (in the classification
of Hiscock and Lindblom \cite{HiscockLindblom}), similar to the Eckart theory
for ordinary fluids. However, it is known that first order theories have stability
and causality problems, and therefore we should expect the same problems will
occur in these simplified reheating scenarios (\cite{HiscockLindblom2}). 

The task at hand is then to write down classical dissipative models of reheating,
in order to account for the basic aspects of the quantum phenomenology, but
to do so in the framework of a consistent relativistic hydrodynamics, thus building
into the model the general principles of relativistic invariance, stability,
causality and the Second Law of Thermodynamics from scratch. 

In an important precedent to this work, for example, Maartens, Pav\'{o}n, Zimdahl
and others \cite{HiscockSamonson,PZ1,PZ2,PZ3,MPZ,CJ,CJP} have developed a dissipative
model of reheating based on the ''truncated'' Israel - Stewart framework.
However, these authors treated the inflaton as a fluid, thus losing the aspects
of the problem associated with the coherence of the inflaton field. Besides,
they focused on the dissipative effects arising from the fact that a mixture
of otherwise ideal fluids will generally develop a nonvanishing bulk viscosity.
Therefore, in their model dissipation happens only if the Universe expands.
On the other hand, quantum fields show dissipative behavior even in Minkowski
space time. In our work, we shall not only treat the inflaton as a coherent
field, but also focus on dissipation arising from having the inflaton not in
equilibrium with the ensemble of all other (quantum) fields. In fact, in this
preliminary investigation we shall work on flat space time backgrounds, although
a fully covariant generalization will be immediate. It is likely that a truly
realistic account of reheating will require a combination of both this and the
Maartens - Pav\'{o}n - Zimdahl approaches. 

Another important difference between this and earlier work is that, instead
of the Israel - Stewart \cite{Israel,IsraelStewart,Maartens} framework (or
similar frameworks, such as ''Extended Thermodynamics'' \cite{EET1}) , we
shall work within the class of ''Divergence Type Theories'' (DTT) earlier
introduced by Geroch \cite{Geroch}. DTTs are appealing because they represent
a mathematically consistent, closed system, rather than a truncated expansion
in deviations from ideal behavior. This makes the discussion of such properties
as causality and stability most transparent \cite{Geroch,Ortiz}, for which
reason DTTs are a natural language to describe the complex nonlinear interactions
between the inflaton field and the fluid of quantum fluctuations during reheating. 

To summarize, in this paper we explore DTTs describing the interaction between
a classical, nonlinear scalar field and a perfect fluid. Our main goal is to
study the constraints on possible interactions which follow from the requirements
of covariance, causality, stability and the Second Law. For simplicity, we shall
work on a flat space - time, and when no loss of generality is involved, we
shall assume homogeneous configurations. The linearized dynamics of the scalar
field around equilibrium is given by the telegraphist's equation, but the characteristic
\( \Gamma \dot{\phi } \) dissipative term appears only as a linear and local
approximation to a more complex term (as it happens in dissipative equations
derived from quantum field theory \cite{CH89,CH97}). 

We shall begin our analysis by constructing a DTT whose solutions, under suitable
restrictions on initial conditions, reduce to solutions of a nonlinear Klein
- Gordon (KG) equation. Let us note that while it is easy to show that the KG
equation is hyperbolic (\cite{Geroch2}), to the best of our knowledge this
equation was never formulated as a DTT. The dynamics of our DTT theory is given
in terms of the conservation equations for the energy momentum tensor and one
vector current \( j_{a} \), and therefore includes one vectorial and one scalar
Lagrange multipliers \( \beta _{\left( c\right) }^{a} \) and \( \xi  \), respectively.
Together with the scalar field \( \phi  \), these are the degrees of freedom
of the theory. 

This is in contradistinction with the usual view of \( \phi  \) and its gradient
\( \phi _{,a} \) as the only degrees of freedom. The extra variable is associated
to dissipation and will become a true dynamical variable only when interaction
with other fluids is introduced. The fact that dissipative theories require
more degrees of freedom than ideal theories is of course rather generic. 

Our next step is to introduce the ensemble of quantum fluctuations of all other
matter fields. For simplicity, we shall describe this as a single ideal fluid.
We shall assume there are no conserved currents other than energy - momentum,
so this fluid will introduce only one new vectorial degree of freedom \( \beta _{(q)}^{a} \),
representing the inverse temperature vector. 

The interacting theory is different from the decoupled theory in two ways. First,
the dependence of the energy momentum tensors for each component (field and
fluid) on the dynamical variables is not the same (in a derivation from first
principles, this would result from taking the variation of interaction and radiative
correction terms in the quantum effective action with respect to the metric
\cite{BirDav}). For simplicity, we shall assume that the energy - momentum
tensors, but not the current \( j^{a} \), are modified in this way. Second,
only the total energy momentum is conserved, so the divergence of the energy
- momentum tensor of the field alone, say, is not zero. We may also modify the
Klein - Gordon equation (near equilibrium, this modification will result in
the addition of a \( \Gamma \dot{\phi } \) dissipative term). As it will turn
out, the possible modifications in the energy momentum tensors are constrained
by causality, while the nonconservation terms in the equations of motion are
constrained by the Second Law. 

After analyzing the constraints derived from linear stability and causality,
we shall conclude this paper by displaying the simplest possible family of theories
satisfying all the physical requirements above. We shall use these theories
for a qualitative discussion of the nonlinear aspects of the approach to equilibrium. 

The rest of the paper is organized as follows. In next section, we show a well
defined DTT which reduces to the KG equation if suitable boundary conditions
are enforced. In Section III we study the mixture of field and fluid in the
DTT framework. We analyze the equilibrium states and their linear stability.
In Section IV we write down a simple, acceptable, nonlinear DTT theory, and
use it to analyze the thermalization process. We summarize the main conclusions
in some brief final remarks. 

We gather in the appendix some elementary facts about DTTs and their connection
to causality.

\section{Divergence type theory of the Klein Gordon field}

To investigate properly a relativistic fluid, the context of DTTs proposed by
Geroch is particularly interesting since causality is easily investigated and/or
achieved. We use signature \( (-,+,+,+) \), latin letters \( a,b,c,... \)
indicate spacetime coordinates and Greek letters \( \mu ,\nu ,... \) indicates
spatial coordinates, other conventions follows Misner, Thorne and Wheeler (\cite{MTW}).
We will also follow the usual (physicist's) convention to designate tensorial
character by its elements, that is \( T^{ab} \) will designate the tensor \( T=T^{ab}e_{a}\otimes e_{b} \),
context indicating easily whether one is making a statement about the whole
tensor or its coordinates. 

Our first goal is to derive a model which contains the classical Klein-Gordon
field as particular case. Since we want to be able to apply this to the early
universe, we allow the classical field to be subject to a non-trivial potential
(which may even be non-renormalizable \cite{CKLL,MTurner,reconstruction}).
The Klein - Gordon theory may be described as a conservation law for the ''current''
\( \phi _{,a} \), namely, \( \nabla ^{a}\phi ,_{a}=V^{\prime }(\phi ). \)
The energy - momentum tensor is also defined in terms of \( \phi _{,a} \) as
\( T_{ab}=\phi _{,a}\phi _{,b}-g_{ab}\left( \phi _{,c}\phi ^{,c}/2+V(\phi )\right)  \).
We therefore postulate that our theory is defined by two currents, energy -
momentum \( T_{ab} \) and a vector current \( j^{a} \), with dynamical equations
\begin{eqnarray}
T_{\quad ;b}^{ab} & = & 0\label{larger1} \\
j_{\quad ;a}^{a} & = & R\left[ x\right] \label{larger2} 
\end{eqnarray}

and the constitutive relation 
\begin{equation}
\label{constitutive}
T^{ab}=j^{a}j^{b}-g^{ab}\left( \frac{1}{2}j_{c}j^{c}+T\left[ R\right] \right) 
\end{equation}

The scalar field \( \phi  \) is introduced by writing the functional relationship
between \( R \) and \( T \) parametrically as \( R=V^{\prime }(\phi ) \),
\( T=V(\phi ). \) This implies no loss of generality. 

Next consider the conservation law 
\begin{equation}
\label{conservation}
T_{\quad ;b}^{ab}=g^{ab}\left( j_{b\, \, ,c}-j_{c\, \, ,b}\right) j^{c}+V^{\prime }\left( j^{a}-\phi ^{,a}\right) =0
\end{equation}

Then if 
\begin{equation}
\label{jab}
j_{b\, \, ,c}-j_{c\, \, ,b}=0
\end{equation}
 we have \( j^{a}=\phi ^{,a} \) and we fall back on the usual (classical) Klein-Gordon
theory. Note that equation (\ref{jab}) is a constraint rather than a dynamical
equation. Defining \( M_{ab}\equiv j_{a;b}-j_{b;a} \) we have the following
identity 
\begin{equation}
j^{c}M_{ab;c}=V^{\prime }M_{ba}+\frac{V^{\prime \prime }}{V^{\prime }}j^{c}\left( j_{b}M_{ac}-j_{a}M_{bc}\right) +j_{\; ;a}^{c}M_{bc}-j_{\; ;b}^{c}M_{ac}
\end{equation}

Therefore if (\ref{jab}) is true initially, it will stay true for all time.
In other words, our set of equations (\ref{larger1}), (\ref{larger2}) and
(\ref{constitutive}) represent a theory larger than Klein-Gordon, reducing
to it if the constraint (\ref{jab}) is enforced initially. 

Our next step is to cast this theory within the DTT framework. Since there are
two currents, we introduce two Lagrange multipliers \( \xi ,\beta _{\left( c\right) }^{a} \)
(we also define \( \beta _{\left( c\right) }=\sqrt{-\beta _{\left( c\right) }^{a}\beta _{\left( c\right) a}} \))
as dynamical degrees of freedom. \( \xi  \) is analogous to a chemical potential
conjugated to the current \( j^{a}, \) while \( \beta _{\left( c\right) }^{a} \)
plays the role of ''inverse temperature'' and is conjugated to \( T^{ab} \).
Observe the perfect-fluid form of the energy-momentum tensor \( T^{ab}. \)

Following the general DTT construction (see Appendix), we introduce the generating
function \( \chi ^{a}=\beta _{\left( c\right) }^{a}p, \) where \( p \) is
the pressure and 
\begin{equation}
\label{Ja}
j^{a}=\beta _{\left( c\right) }^{a}\frac{\partial p}{\partial \xi }
\end{equation}
\begin{equation}
T^{ab}=pg^{ab}-\beta _{\left( c\right) }^{a}\beta _{\left( c\right) }^{b}\frac{1}{\beta _{\left( c\right) }}\frac{\partial p}{\partial \beta _{\left( c\right) }}=pg^{ab}-j^{a}j^{b}\frac{1}{\beta _{\left( c\right) }}\frac{\left( \frac{\partial p}{\partial \beta _{\left( c\right) }}\right) }{\left( \frac{\partial p}{\partial \xi }\right) ^{2}}
\end{equation}

leading to the following equation upon comparison with (\ref{constitutive})
\begin{eqnarray}
-\frac{1}{\beta _{\left( c\right) }}\frac{\partial p}{\partial \beta _{\left( c\right) }} & = & \left( \frac{\partial p}{\partial \xi }\right) ^{2}\nonumber \\
p & = & -\frac{1}{2}j_{c}j^{c}-T
\end{eqnarray}

The first one is a differential equation in the variables \( \xi ,\beta _{\left( c\right) } \)
with solution 
\begin{equation}
p=\frac{\xi ^{2}}{2\beta _{\left( c\right) }^{2}}+\Lambda 
\end{equation}

where \( \Lambda  \) is independent of \( \xi  \) and \( \beta _{\left( c\right) } \)
but can depend on spacetime coordinates. In fact, using the second equation
we see that 
\begin{equation}
\Lambda =-V(\phi )
\end{equation}

Thus 
\begin{equation}
\label{Ja2}
j^{a}=\frac{\xi }{\beta _{\left( c\right) }^{2}}\beta _{\left( c\right) }^{a}
\end{equation}

Recall that the conservation law for \( T^{ab} \), eq. (\ref{conservation})
implies \( j^{2}-j^{a}\phi _{,a}=0 \) even when the constraint eq. (\ref{jab})
is not enforced. Since \( j^{2}=-(\xi /\beta _{\left( c\right) })^{2} \), we
get 
\begin{equation}
\label{xidef}
\beta _{\left( c\right) }^{a}\phi _{,a}=-\xi 
\end{equation}

Finally, we easily compute \( \chi _{c} \) using 
\begin{equation}
p=-\frac{1}{\beta _{\left( c\right) }}\frac{\partial \chi _{c}}{\partial \beta _{\left( c\right) }}
\end{equation}

Leading immediately to 

\begin{equation}
\chi _{c}=-\frac{1}{2}\xi ^{2}\ln \beta _{\left( c\right) }+\frac{1}{2}T\beta _{\left( c\right) }^{2}
\end{equation}

We can regard eq. (\ref{xidef}) as a generalization of the canonical momentum
\( \pi =\dot{\phi } \); in the ''rest'' frame where \( \beta _{\left( c\right) }^{a}=\left( \beta _{\left( c\right) },\vec{0}\right) , \)
we get \( \pi =-\xi /\beta _{\left( c\right) }. \) The problem is that only
this ratio has a direct meaning in terms of the Klein - Gordon theory \textit{alone},
that is, the KG equations are invariant under a rescaling of \( \xi  \) and
\( \beta _{\left( c\right) }^{a} \) by a common factor. In order to break this
indeterminacy, we must look at the larger framework where the Klein - Gordon
field interacts with other fluids. Then we complete the definition of \( \xi  \)
and \( \beta _{\left( c\right) }^{a} \) by demanding that, in equilibrium,
\( \beta _{\left( c\right) }^{a} \) must be identical to the (only) inverse
temperature vector of the full theory. 

Knowing \( \beta _{\left( c\right) }^{a} \), we now regard eq. (\ref{xidef})
as the \textit{definition} of the scalar \( \xi  \). This means that, while
for the pure KG theory eq. (\ref{xidef}) simply follows from energy - momentum
conservation, we shall demand it also holds unchanged in the interacting theory.
This procedure is of course suggested by Landau and Lifschitz' treatment of
the damped harmonic oscillator in ref. \cite{LL}

Since \( \chi _{c}^{a} \) is a homogeneous function of degree \( 1 \) in \( \xi  \)
and \( \beta _{\left( c\right) }^{a} \), the entropy current and entropy creation
rate vanish in the pure KG theory, as expected for a coherent field.

\section{Divergence type theory of interacting fields and fluids}

Our next goal is to describe the interaction between the KG field and other
forms of matter in the context of DTTs. We thus introduce a (perfect) fluid,
described by an inverse temperature vector \( \beta _{(q)}^{a} \) and energy
- momentum tensor \( T_{q}^{ab} \) derivable from a generating functional \( \chi _{q}^{a}=\partial \chi _{q}/\partial \beta _{(q)\, a} \)
(see the Appendix). The interacting theory shall be described by the equations
\begin{eqnarray}
j_{\; ;a}^{a} & = & R+\Delta \nonumber \\
T_{c\; \; ;b}^{ab} & = & I^{a}\nonumber \\
T_{q\; \; ;b}^{ab} & = & -I^{a}\nonumber \\
\beta _{(c)}^{a}\phi _{,a} & = & -\xi \label{system} 
\end{eqnarray}

Where \( j^{a} \) and \( T_{c}^{ab} \) are the current and energy - momentum
tensor for the inflaton field, and we have added the definition eq. (\ref{xidef}),
which, unlike the situation in the noninteracting theory, is now independent
of the other equations . Let us seek a generating functional of the form

\begin{equation}
\chi ^{a}=\chi _{c}^{a}+\chi _{q}^{a}+\Xi ^{a}
\end{equation}

The total system is generated not only by the sum of each thermodynamic potential
but there is a third potential to include the interaction between field and
fluid. Each energy - momentum tensor will be given by: 
\begin{eqnarray}
T_{c}^{ab} & = & \frac{\partial \chi _{c}^{a}}{\partial \beta _{(c)\, b}}+\frac{\partial \Xi ^{a}}{\partial \beta _{(c)\, b}}\nonumber \\
T_{q}^{ab} & = & \frac{\partial \chi _{q}^{a}}{\partial \beta _{(q)\, b}}+\frac{\partial \Xi ^{a}}{\partial \beta _{(q)\, b}}
\end{eqnarray}

Let's define the following variables 
\begin{eqnarray}
\beta ^{a} & = & \frac{1}{2}\left( \beta _{(c)}^{a}+\beta _{(q)}^{a}\right) \nonumber \\
B^{a} & = & \beta _{(c)}^{a}-\beta _{(q)}^{a}
\end{eqnarray}

Then 
\begin{eqnarray}
\frac{\partial \Xi ^{a}}{\partial \beta _{(c)\, b}} & = & \frac{1}{2}\frac{\partial \Xi ^{a}}{\partial \beta _{b}}+\frac{\partial \Xi ^{a}}{\partial B_{b}}\nonumber \\
\frac{\partial \Xi ^{a}}{\partial \beta _{(q)\, b}} & = & \frac{1}{2}\frac{\partial \Xi ^{a}}{\partial \beta _{b}}-\frac{\partial \Xi ^{a}}{\partial B_{b}}
\end{eqnarray}

Thus 
\begin{equation}
T_{c}^{ab}+T_{q}^{ab}=\frac{\partial \chi _{c}^{a}}{\partial \beta _{(c)\, b}}+\frac{\partial \chi _{q}^{a}}{\partial \beta _{(q)\, b}}+\frac{\partial \Xi ^{a}}{\partial \beta _{b}}
\end{equation}

Note that the only real physical system is the one described by the total energy-momentum
tensor given above. Any separation in two fluids will be tinged with arbitrariness,
a fact that will be clear soon. We will ask that the total energy-momentum tensor
be symmetric; therefore 
\begin{equation}
\Xi ^{a}=\frac{\partial \Xi }{\partial \beta _{a}}
\end{equation}

\( \Xi  \) will depend in general on scalars as demanded by Lorentz invariance;
namely \( \Xi =\Xi (\xi ,u,v,w,\phi ) \) where 
\begin{eqnarray}
u & = & -\beta _{a}\beta ^{a}\nonumber \\
v & = & -B_{a}B^{a}\nonumber \\
w & = & -\beta _{a}B^{a}
\end{eqnarray}

Then 
\begin{eqnarray}
\frac{\partial }{\partial \beta _{a}} & = & -2\beta ^{a}\frac{\partial }{\partial u}-B^{a}\frac{\partial }{\partial w}\nonumber \\
\frac{\partial }{\partial B_{a}} & = & -2B^{a}\frac{\partial }{\partial v}-\beta ^{a}\frac{\partial }{\partial w}
\end{eqnarray}

Note that even if the source term \( I^{b} \) is taken to be null, the energy
momentum tensors for field and fluid do not fall back automatically to their
old form. Since the only ``true'' energy-momentum tensor is the total one,
it is sometimes helpful and instructive to rewrite the equation of motions of
the \( T_{i}^{ab} \) (\( i=c,q \)). Let us define the following 
\begin{eqnarray}
T_{+}^{ab} & = & T_{c}^{ab}+T_{q}^{ab}\nonumber \\
T_{-}^{ab} & = & T_{c}^{ab}-T_{q}^{ab}
\end{eqnarray}

Note that

\begin{equation}
T_{+}^{ab}=\frac{\partial \chi ^{a}}{\partial \beta _{b}}
\end{equation}

since \( \partial \chi _{q}^{a}/\partial \beta _{(c)\, b}=0=\partial \chi _{c}^{a}/\partial \beta _{(q)\, b} \).
Also 
\begin{equation}
T_{-}^{ab}=2\frac{\partial \chi ^{a}}{\partial B_{b}}
\end{equation}

The following identities follow straightforwardly 
\begin{equation}
\label{chi_{a}}
\Xi ^{a}=-2\beta ^{a}\frac{\partial \Xi }{\partial u}-B^{a}\frac{\partial \Xi }{\partial w}
\end{equation}

\begin{equation}
\label{chi_{a}_{b}}
\frac{\partial \Xi ^{a}}{\partial \beta _{b}}=-2g^{ab}\frac{\partial \Xi }{\partial u}+4\beta ^{a}\beta ^{b}\frac{\partial ^{2}\Xi }{\partial u^{2}}+B^{a}B^{b}\frac{\partial ^{2}\Xi }{\partial w^{2}}+2\left( \beta ^{a}B^{b}+B^{a}\beta ^{b}\right) \frac{\partial ^{2}\Xi }{\partial u\partial w}
\end{equation}

\begin{equation}
\label{chi_{a}b_{m}enos}
\frac{\partial \Xi ^{a}}{\partial B_{b}}=-g^{ab}\frac{\partial \Xi }{\partial w}+2\beta ^{a}\beta ^{b}\frac{\partial ^{2}\Xi }{\partial u\partial w}+2B^{a}B^{b}\frac{\partial ^{2}\Xi }{\partial v\partial w}+4\beta ^{a}B^{b}\frac{\partial ^{2}\Xi }{\partial u\partial v}+B^{a}\beta ^{b}\frac{\partial ^{2}\Xi }{\partial w^{2}}
\end{equation}

Let 's compute the entropy creation rate. Using 
\begin{eqnarray}
\frac{\partial }{\partial \beta _{b}} & = & \frac{\partial }{\partial \beta _{(c)\, b}}+\frac{\partial }{\partial \beta _{(q)\, b}}\nonumber \\
\frac{\partial }{\partial B_{b}} & = & \frac{1}{2}\frac{\partial }{\partial \beta _{(c)\, b}}-\frac{1}{2}\frac{\partial }{\partial \beta _{(q)\, b}}
\end{eqnarray}

we find 
\begin{equation}
\nabla _{a}S^{a}=\frac{\partial \Xi ^{a}}{\partial \phi }\nabla _{a}\phi -B_{b}I^{b}-\xi \Delta 
\end{equation}

where we used the fact that \( \beta _{(c)}^{a}\phi _{,a}+\xi =0. \) For simplicity,
let us ask that \( \Xi  \) be independent of \( \phi , \) so that the entropy
production reduces to 
\begin{equation}
\nabla _{a}S^{a}=-B_{b}I^{b}-\xi \Delta 
\end{equation}

The second law of thermodynamics imposes that \( \nabla _{a}S^{a}>0. \) This
means that \( I^{b} \) and \( \Delta  \) must vanish when \( B^{b} \) and
\( \xi  \) go to zero. In this limit, \( \Delta , \) which is a scalar, must
take the form \( \Delta =A\xi +Bw \) (\( v \) being of higher order); Lorentz
invariance demands \( I^{b}=-\left( C\xi +Dw\right) \beta ^{b}-EB^{b}. \) Therefore
\( \nabla _{a}S^{a}=-A\xi ^{2}-\left( B+C\right) w\xi -Dw^{2}-Ev, \) and we
must have \( A,D\leq 0, \) \( E\leq 0 \) and \( AD\geq \left( B+C\right) ^{2}/4. \)
Equivalently, we may parametrize \( A=M^{\xi \xi } \), \( B\beta ^{a}=2(1-\kappa )M^{\xi a} \),
\( C\beta ^{b}=2\kappa M^{\xi b} \) and \( D\beta ^{a}\beta ^{b}+Eg^{ab}=M^{ab} \),
whereby 
\begin{eqnarray}
\Delta  & \equiv  & M^{\xi \xi }\xi +2B_{a}(1-\kappa )M^{\xi a}\label{matrix} \\
I^{b} & \equiv  & 2\kappa \xi M^{\xi b}+B_{a}M^{ab}\nonumber 
\end{eqnarray}

\subsection{First order analysis away from equilibrium in a simplified, homogeneous model}

Let 's turn now to the equation of motion to analyze small deviations from equilibrium.
The requirement that the equilibrium state must be isotropic in some ''rest''
frame implies that, in equilibrium, the two vector \( \beta _{(c)}^{a} \) and
\( \beta _{(q)}^{a} \) must be parallel. Therefore we can write \( \beta _{(c)}^{a}=\beta _{c}u^{a} \)
and \( \beta _{(q)}^{a}=\beta _{q}u^{a}, \) with a common unit vector \( u^{a} \).
We will also restrict ourselves to the homogeneous case. Observe that \( T_{c\; ;b}^{\mu b}=T_{q\; ;b}^{\mu b}=0 \)
identically for \( \mu \neq 0, \) so there are only four nontrivial equations,
including eq. (\ref{xidef}).

In equilibrium, we must have \( \xi =\Delta =I^{a}=R=0. \) Defining \( \phi =0 \)
at the equilibrium point, we conclude \( \left( \xi _{eq\, },\beta _{eq\, },B_{eq\, },\phi _{eq}\right) =\left( 0,\beta _{0},0,0\right)  \).
We can now analyze the first order deviations away from equilibrium; that is
\begin{eqnarray}
\xi  & = & \delta \xi \nonumber \\
\phi  & = & \delta \phi \nonumber \\
\beta  & = & \beta _{eq}+\delta \beta \nonumber \\
B & = & \delta B
\end{eqnarray}

We can expand

\begin{equation}
\Xi =\sum _{n=0}^{\infty }F_{n}\left( \beta ,\xi \right) B^{n+2}
\end{equation}

\( \Xi  \) is constrained to this form by demanding that at equilibrium, where
\( \beta _{c}=\beta _{q} \), we fall back to our initial \( T^{ab} \) and
\( j^{a}. \) \( \Delta  \) and \( I^{b} \) are given by eq. (\ref{matrix}),
where all the matrix element are evaluated at the equilibrium value \( \xi _{eq}= \)
\( B_{eq}^{a}=0, \) \( \beta _{eq}^{a}=\beta _{0}^{eq} \) and \( \phi _{0}=0. \)
Moreover we will define \( m^{2}=V^{\prime \prime }\left( 0\right)  \), so
to first order 
\begin{equation}
R=m^{2}\delta \phi 
\end{equation}

In our case we have 
\begin{eqnarray}
j_{\; \; ;a}^{a} & = & \frac{1}{\beta _{0}}\left( \delta \xi \right) _{,t}=M_{0}^{\xi \xi }\delta \xi +2(1-\kappa )M_{0}^{\xi 0}\delta B+m^{2}\delta \phi \nonumber \\
T_{c\; ;b}^{0b} & = & -\frac{\partial ^{3}\Xi }{\partial \beta \partial B^{2}}\left( \delta B\right) _{,t}=2\kappa M_{0}^{\xi 0}\delta \xi +M_{0}^{00}\delta B\nonumber \\
T_{q\; ;b}^{0b} & = & \rho _{q}^{\prime }\left( \delta \beta \right) _{,t}+\left( \frac{\partial ^{3}\Xi }{\partial \beta \partial B^{2}}-\frac{1}{2}\rho _{q}^{\prime }\right) \left( \delta B\right) _{,t}=-2\kappa M_{0}^{\xi 0}\delta \xi -M_{0}^{00}\delta B\nonumber \\
\beta _{0}\left( \delta \phi \right) _{,t} & = & -\delta \xi 
\end{eqnarray}

We can revert to our old coordinates \( \delta \beta _{c} \) and \( \delta \beta _{q} \)

\[
\frac{1}{\beta _{0}}\left( \delta \xi \right) _{,t}=M_{0}^{\xi \xi }\delta \xi +2(1-\kappa )M_{0}^{\xi 0}\delta \beta _{c}+2(1-\kappa )M_{0}^{\xi 0}\delta \beta _{q}+m^{2}\delta \phi \]

\begin{eqnarray}
-\frac{\partial ^{3}\Xi }{\partial \beta \partial B^{2}}\left( \delta \beta _{c}\right) _{,t}+\frac{\partial ^{3}\Xi }{\partial \beta \partial B^{2}}\left( \delta \beta _{q}\right) _{,t} & = & 2\kappa M_{0}^{\xi 0}\delta \xi +M_{0}^{00}\delta \beta _{c}-M_{0}^{00}\delta \beta _{q}\nonumber \\
\left( \frac{\partial ^{3}\Xi }{\partial \beta \partial B^{2}}\right) \left( \delta \beta c\right) _{,t}+\left( \rho _{q}^{\prime }-\frac{\partial ^{3}\Xi }{\partial \beta \partial B^{2}}\right) \left( \delta \beta _{q}\right) _{,t} & = & -2\kappa M_{0}^{\xi 0}\delta \xi -M_{0}^{00}\delta \beta _{c}+M_{0}^{00}\delta \beta _{q}\nonumber \\
\beta _{0}\left( \delta \phi \right) _{,t} & = & -\delta \xi 
\end{eqnarray}

The advantage to do this is that we can add the middle two to obtain 
\begin{equation}
\rho _{q}^{\prime }\left( \delta \beta _{q}\right) _{,t}=0
\end{equation}

That is 
\begin{equation}
\delta \beta _{q}=constant
\end{equation}

We are thus left with 
\begin{eqnarray}
\frac{1}{\beta _{0}}\left( \delta \xi \right) _{,t} & = & M_{0}^{\xi \xi }\delta \xi +2(1-\kappa )M_{0}^{\xi 0}\delta \beta _{c}+m^{2}\delta \phi \nonumber \\
-\frac{\partial ^{3}\Xi }{\partial \beta \partial B^{2}}\left( \delta \beta _{c}\right) _{,t} & = & 2\kappa M_{0}^{\xi 0}\delta \xi +M_{0}^{00}\delta \beta _{c}\nonumber \\
\beta _{0}\left( \delta \phi \right) _{,t} & = & -\delta \xi 
\end{eqnarray}

or, in matrix form 
\begin{equation}
\label{equations}
\left( \begin{array}{lll}
\left( \beta _{0}\right) ^{-1} & 0 & 0\\
0 & -\frac{\partial ^{3}\Xi }{\partial \beta \partial B^{2}} & 0\\
0 & 0 & \beta _{0}
\end{array}\right) \left( \begin{array}{l}
\delta \xi \\
\delta \beta _{c}\\
\delta \phi 
\end{array}\right) _{,t}=\left( \begin{array}{lll}
M_{0}^{\xi \xi } & 2(1-\kappa )M_{0}^{\xi 0} & m^{2}\\
2\kappa M_{0}^{\xi 0} & M_{0}^{00} & 0\\
-1 & 0 & 0
\end{array}\right) \left( \begin{array}{l}
\delta \xi \\
\delta \beta _{c}\\
\delta \phi 
\end{array}\right) 
\end{equation}

We will consider solutions of the type \( \exp \left( -\gamma t\right) . \)
There is a solution if 
\begin{equation}
\left| \begin{array}{lll}
\left( -\gamma \beta _{0}^{-1}-M_{0}^{\xi \xi }\right)  & -2(1-\kappa )M_{0}^{\xi 0} & -m^{2}\\
-2\kappa M_{0}^{\xi 0} & \left( \gamma \frac{\partial ^{3}\Xi }{\partial \beta \partial B^{2}}-M_{0}^{00}\right)  & 0\\
1 & 0 & -\gamma \beta _{0}
\end{array}\right| =0
\end{equation}

To make notation less clumsy, let us write 
\begin{eqnarray}
M_{0}^{\xi \xi } & = & \varepsilon _{1}=A\nonumber \\
\frac{\partial ^{3}\Xi }{\partial \beta \partial B^{2}} & = & X\nonumber \\
M_{0}^{\xi 0} & = & \varepsilon _{2}=\left( B+C\right) \frac{\beta ^{0}}{2}\nonumber \\
M_{0}^{00} & = & \varepsilon _{3}=D\beta _{0}^{2}+E\label{values} 
\end{eqnarray}

The notation is chosen to emphasize explicitly which quantities are small and
which aren't . Namely, the \( \varepsilon _{i} \) are (very) small but the
\( X \) is not. Therefore 
\begin{equation}
\left| \begin{array}{lll}
\left( -\gamma \beta _{0}^{-1}-\varepsilon _{1}\right)  & -2\left( 1-\kappa \right) \varepsilon _{2} & -m^{2}\\
-2\kappa \varepsilon _{2} & \left( \gamma X-\varepsilon _{3}\right)  & 0\\
1 & 0 & -\gamma \beta _{0}
\end{array}\right| =0
\end{equation}

Expanding with the second column we have 
\begin{equation}
4\kappa \left( 1-\kappa \right) \gamma \beta _{0}\varepsilon _{2}^{2}+X\beta _{0}\gamma ^{2}\varepsilon _{1}-\beta _{0}\gamma \varepsilon _{1}\varepsilon _{3}+X\gamma \left( \gamma ^{2}+m^{2}\right) -\varepsilon _{3}\left( \gamma ^{2}+m^{2}\right) =0
\end{equation}

The solutions to order zero are \( \gamma =\pm im \) and \( \gamma =0 \).
To order one, the equation reduces to 
\begin{equation}
X\beta _{0}\gamma ^{2}\varepsilon _{1}+X\gamma \left( \gamma ^{2}+m^{2}\right) -\varepsilon _{3}\left( \gamma ^{2}+m^{2}\right) =0
\end{equation}

Writing first \( \gamma =\pm im+\delta  \) we obtain 
\begin{equation}
\delta =-\frac{\beta _{0}\varepsilon _{1}}{2}
\end{equation}

and in the case \( \gamma =\delta  \) we have 
\begin{equation}
\delta =\frac{\varepsilon _{3}}{X}
\end{equation}

That is 
\begin{eqnarray}
\gamma  & = & \pm im-\frac{\beta _{0}\varepsilon _{1}}{2}\nonumber \\
\gamma  & = & \frac{\varepsilon _{3}}{X}
\end{eqnarray}

Notice that \( \varepsilon _{1}<0 \) and \( \varepsilon _{3}<0 \) . Since
\( \beta _{0}>0 \) then the first two solutions corresponding to oscillatory
modes get a damping part. The other solution will also be damped if 
\begin{equation}
X\equiv \frac{\partial ^{3}\Xi }{\partial \beta \partial B^{2}}<0
\end{equation}

In this case the system will be causal if the matrix \( M^{0AB}v \) is negative-definite
(where \( v=v^{0}>0 \) and \( v_{0}=-v^{0} \), since it is the temporal component
of a temporal 4-vector oriented toward the future) and where 
\[
M^{0AB}v=\left( \begin{array}{lll}
-v\beta _{0}^{-1} & 0 & 0\\
0 & v\frac{\partial ^{3}\Xi }{\partial \beta \partial B^{2}} & 0\\
0 & 0 & -v\beta _{0}
\end{array}\right) \]

Obviously, this matrix is negative-definite if 
\begin{eqnarray*}
\beta _{0} & > & 0\\
\frac{\partial ^{3}\Xi }{\partial \beta \partial B^{2}} & < & 0
\end{eqnarray*}

We thus verify that the causality condition implies the stability of the equilibrium
solution. This is in agreement with the findings of Hiscock and Lindblom in
the context of the Israel - Stewart formalism \cite{HiscockLindblom}.

Going one order further we find 
\begin{eqnarray}
\gamma  & = & \pm i\left\{ m-\frac{\beta _{0}}{2m}\left[ \frac{1}{4}\beta _{0}\varepsilon _{1}^{2}-\frac{4\kappa \left( 1-\kappa \right) }{X}\varepsilon _{2}^{2}\right] \right\} -\frac{\beta _{0}\varepsilon _{1}}{2}\label{complex} \\
\gamma  & = & \frac{\varepsilon _{3}}{X}\label{real} 
\end{eqnarray}

That is the correction to the third solution is at least third order and the
second order terms modify the frequency of the oscillation in the case of the
other two solutions.

If we look back to the linearized equation of motion for \( \delta \phi , \)
we obtain

\begin{equation}
-\left( \delta \phi \right) _{,tt}=-M_{0}^{\xi \xi }\beta _{0}\left( \delta \phi \right) _{,t}+2(1-\kappa )M_{0}^{\xi 0}\delta \beta _{c}+m^{2}\delta \phi 
\end{equation}

In the simplest case \( M_{0}^{\xi 0}=0 \), this is the telegraphist equation
with a damping term \( \Gamma \dot{\phi } \), where \( \Gamma =\left| M_{0}^{\xi \xi }\right| \beta _{0}. \)
It is important to realize, however, that this identification holds only to
linear order away from equilibrium. In general, \( \Gamma  \) will not be a
constant, but a function of the dynamical variables \( \beta _{q} \) and \( \beta _{c}, \)
and therefore it will depend on the history of the system.

We may observe that when \( M_{0}^{\xi 0}=0, \) we get \( \beta _{c},_{t}=0 \)
to first order.

\section{A simple nonlinear model of field - fluid interaction}

In this section we shall investigate the simplest DTTs of field - fluid interaction
satisfying the requirements of causality, stability and the Second Law discussed
in the previous Section. To make things simplest, we will restrict the form
of \( \Xi  \) in the following manner 
\begin{equation}
\Xi =F(u)v+G(u)w^{2}
\end{equation}

and we will investigate what restrictions one can expect from causality on \( F \)
and \( G \). By making \( \Xi  \) independent of \( \xi , \) we make sure
that the current \( j^{a} \) preserves its form in the interacting theory.
We also assume \( \Xi  \) to be \( \phi - \) independent, and require it to
be quadratic on the difference variable \( B^{a} \), in such a way that corrections
to the energy - momentum tensors will vanish in the equilibrium state. Observe
that 
\[
\Xi ^{a}=-2\beta ^{a}\left( F^{\prime }v+G^{\prime }w^{2}\right) -2B^{a}Gw\]

\begin{eqnarray*}
\frac{\partial \Xi ^{a}}{\partial \beta _{b}} & = & 4\beta ^{a}\beta ^{b}\left( F^{\prime \prime }v+G^{\prime \prime }w^{2}\right) -2g^{ab}\left( F^{\prime }v+G^{\prime }w^{2}\right) \\
 &  & +4\left( \beta ^{a}B^{b}+B^{a}\beta ^{b}\right) G^{\prime }w+2B^{a}B^{b}G
\end{eqnarray*}

\begin{equation}
\frac{\partial \Xi ^{a}}{\partial B_{b}}=4\beta ^{a}\beta ^{b}G^{\prime }w-2g^{ab}Gw+4\beta ^{a}B^{b}F^{\prime }+2B^{a}\beta ^{b}G
\end{equation}

At equilibrium 
\begin{equation}
\frac{\partial ^{2}\Xi ^{a}}{\partial \beta _{c}\partial \beta _{b}}=\frac{\partial ^{2}\Xi ^{a}}{\partial B_{c}\partial \beta _{b}}=0
\end{equation}

and 
\begin{equation}
\frac{\partial ^{2}\Xi ^{a}}{\partial B_{c}\partial B_{b}}=-4\beta ^{a}_{(eq)}\beta ^{b}_{(eq)}\beta ^{c}_{(eq)}G^{\prime }+2\left( g^{ab}\beta ^{c}_{(eq)}+g^{ac}\beta ^{b}_{(eq)}\right) G+4\beta ^{a}_{(eq)}g^{bc}F^{\prime }
\end{equation}

Causality demands that

\begin{equation}
\left( \begin{array}{cccc}
\frac{\beta _{(eq)}^{a}v_{a}}{\beta _{\left( c\right) }^{2}} & 0 & 0 & 0\\
0 & \frac{\partial \chi _{c}^{a}}{\partial \beta _{c}\partial \beta _{b}}v_{a} & 0 & 0\\
0 & 0 & \frac{\partial \left( \chi _{q}^{a}+\Xi ^{a}\right) }{\partial B_{c}\partial B_{b}}v_{a} & 0\\
0 & 0 & 0 & \beta _{(eq)}^{a}v_{a}
\end{array}\right) 
\end{equation}

should be negative-definite for any future-oriented, timelike vector \( v^{a} \)
. Since \( \chi _{q}^{a} \) represents a perfect fluid, we obtain causality
under the usual conditions \cite{Geroch}. We only need to verify that \( \frac{\partial \Xi ^{a}}{\partial B_{c}\partial B_{b}}v_{a} \)
is negative-definite by itself. Specifically, we want to know if 
\begin{equation}
\label{quadratic}
\frac{\partial \Xi ^{a}}{\partial B_{c}\partial B_{b}}v_{a}=4\beta _{(eq)}^{a}v_{a}\left( -\beta _{(eq)}^{b}\beta _{(eq)}^{c}G^{\prime }+g^{bc}F\prime \right) +2\left( v^{b}\beta _{(eq)}^{c}+v^{c}\beta _{(eq)}^{b}\right) G
\end{equation}

is negative-definite, that is if \( L_{b}L_{c}\frac{\partial \left( \Xi ^{a}\right) }{\partial B_{c}\partial B_{b}}v_{a}<0 \)
for any \( L_{a}. \) To achieve this, let's decompose \( L_{b} \) into its
longitudinal and transverse part relative to \( \beta _{(eq)\, b} \) : 
\begin{equation}
L_{b}=l\, \left( \frac{\beta _{(eq)\, b}}{u}\right) +R_{b}\qquad ;\qquad R^{b}\beta _{(eq)\, b}=0
\end{equation}

We easily work out the following 
\begin{equation}
4\beta _{(eq)}^{a}v_{a}\left( -G^{\prime }-\frac{\left( F^{\prime }+G\right) }{u}\right) \, l^{2}+4\beta _{(eq)}^{a}v_{a}R_{b}R^{b}F^{\prime }-4R_{b}v^{b}G\, l
\end{equation}

We can already extract some information. Taking \( l=0 \) we have 
\begin{equation}
\beta _{(eq)}^{a}v_{a}R_{b}R^{b}F^{\prime }<0
\end{equation}

which imply that \( F^{\prime }>0 \) since \( \beta _{(eq)}^{a}v_{a}<0 \)
and \( R_{b}R^{b}>0 \) . Also, Taking \( R_{b}=0 \) we obtain 
\begin{equation}
\beta _{(eq)}^{a}v_{a}\left( -G^{\prime }-\frac{\left( F^{\prime }+G\right) }{u}\right) \, l^{2}<0
\end{equation}

this time implying that \( \left( -G^{\prime }-\frac{\left( F^{\prime }+G\right) }{u}\right) >0 \).
Let us now write 
\begin{equation}
v^{b}=\lambda \beta _{(eq)}^{b}+\omega ^{b}\qquad ;\qquad \beta _{(eq)\, b}\omega ^{b}=0
\end{equation}

and decompose the spatial vector \( R_{b} \) into the part which is longitudinal
and transversal to \( \omega ^{b} \): 
\begin{equation}
R^{b}=\eta \omega ^{b}+t^{b}\qquad ;\qquad t_{b}\beta ^{b}_{(eq)}=0=\omega _{b}t^{b}
\end{equation}

Therefore \( R_{b}v^{b}=\eta \omega _{b}\omega ^{b} \) and, replacing in (\ref{quadratic}),
we obtain upon division by \( \beta _{(eq)}^{a}v_{a}<0 \): 
\begin{equation}
\left( -G^{\prime }-\frac{\left( F^{\prime }+G\right) }{u}\right) \, l^{2}-\frac{\omega _{b}\omega ^{b}}{\xi ^{a}v_{a}}G\eta l+\left( \eta ^{2}\omega _{b}\omega ^{b}+t_{b}t^{b}\right) F^{\prime }
\end{equation}

which should be now positive-definite in order to have (\ref{quadratic}) negative-definite.
Since \( t_{b}t^{b}F^{\prime }>0 \) it suffice then to ask 
\begin{equation}
\left( -G^{\prime }-\frac{\left( F^{\prime }+G\right) }{u}\right) \, l^{2}-\frac{\omega _{b}\omega ^{b}}{\beta _{(eq)}^{a}v_{a}}G\eta l+\eta ^{2}\omega _{b}\omega ^{b}F^{\prime }>0
\end{equation}

This is a quadratic form in \( l \) and \( \eta  \) and we are thus demanding
that the matrix 
\begin{equation}
\label{dos_{X}_{d}os}
\left( \begin{array}{cc}
\left( -G^{\prime }-\frac{\left( F^{\prime }+G\right) }{u}\right)  & -\frac{\omega _{b}\omega ^{b}}{2\beta _{(eq)}^{a}v_{a}}G\\
-\frac{\omega _{b}\omega ^{b}}{2\beta _{(eq)}^{a}v_{a}}G & \omega _{b}\omega ^{b}F^{\prime }
\end{array}\right) 
\end{equation}

should be positive-definite. We already know that this means that the diagonal
element are positive, a fact that we already deduced. Now since the matrix is
real and symmetric, we know that its eigenvalues are real. Since the diagonal
elements are positive, then it is sufficient to prove that the determinant is
positive. The determinant is given by 
\begin{equation}
\label{determinant}
\omega _{b}\omega ^{b}\left[ \left( -G^{\prime }-\frac{\left( F^{\prime }+G\right) }{u}\right) F^{\prime }-\frac{\omega _{b}\omega ^{b}}{4\left( \beta _{(eq)}^{a}v_{a}\right) ^{2}}G^{2}\right] >0
\end{equation}

This condition can be simplified by considering that \( v_{a}v^{a}=\lambda ^{2}\beta _{(eq)\, a}\beta _{(eq)}^{a}+\omega _{a}\omega ^{a}<0 \)
that is, using \( \beta _{(eq)\, a}\beta _{(eq)}^{a}=-u \)
\begin{equation}
\omega _{a}\omega ^{a}<\lambda ^{2}u
\end{equation}

Also \( \beta _{(eq)}^{a}v_{a}=-\lambda u \) . Therefore 
\begin{equation}
\frac{\omega _{a}\omega ^{a}}{\left( \beta _{(eq)}^{a}v_{a}\right) ^{2}}=\frac{\omega _{a}\omega ^{a}}{\lambda ^{2}\mu ^{4}}<\frac{1}{u}
\end{equation}

Then if 
\begin{equation}
\left( -G^{\prime }-\frac{\left( F^{\prime }+G\right) }{u}\right) F^{\prime }>\frac{G^{2}}{4}\frac{1}{u}
\end{equation}

then condition (\ref{determinant}) is fulfilled. That is, the matrix (\ref{dos_X_dos})
is positive definite if, 
\begin{equation}
\label{condition_{1}}
F^{\prime }>0
\end{equation}

\begin{equation}
\label{condition_{2}}
\left( G^{\prime }+\frac{\left( F^{\prime }+G\right) }{u}\right) <0
\end{equation}

and 
\begin{equation}
\label{condition_{3}}
\left( G^{\prime }+\frac{\left( F^{\prime }+G\right) }{u}\right) F^{\prime }<-\frac{G^{2}}{4u}
\end{equation}

To see the meaning of these conditions, consider the case where \( F \) and
\( G \) follow power laws. Dimensional analysis in natural units (\( c=\hbar =k_{B}=1) \)
indicates that 
\begin{equation}
[\Xi ]=T^{2}
\end{equation}

Thus, writing \( F=F_{0}u^{\alpha +1} \) and \( G=G_{0}u^{\alpha } \) implies
that \( \alpha =-3 \) , that is 
\begin{equation}
\Xi =F_{0}u^{-2}v+G_{0}u^{-3}w^{2}
\end{equation}

The three conditions (\ref{condition_1}),(\ref{condition_2}) and (\ref{condition_3})
read 
\[
F_{0}<0\]

\[
G_{0}+F_{0}>0\; \Rightarrow G_{0}>-F_{0}>0\]

\begin{equation}
4F_{0}\left( G_{0}+F_{0}\right) <-\frac{1}{4}G_{0}^{2}
\end{equation}

Substituting \( F_{0}/G_{0}\equiv -x \), we have the following inequalities
for \( x \) : 
\begin{equation}
4x^{2}-4x+\frac{1}{4}<0
\end{equation}

implying that \( 1/2-\sqrt{3}/4<x<1/2+\sqrt{3}/4. \) Observe that both limit
cases \( x=0 \) and \( x=1 \) are excluded.

We thus obtain the following final form for \( \Xi  \)
\begin{equation}
\Xi =\frac{\Omega _{0}}{u}\left[ x\left( \frac{v}{u}\right) -\left( \frac{w}{u}\right) ^{2}\right] 
\end{equation}

(\( \Omega _{0}=-G_{0}<0 \)) which can be rewritten in a covariant manner as
(writing \( \mu \equiv \sqrt{-\beta _{a}\beta ^{a}} \)) 
\begin{equation}
\Xi =\frac{\Omega _{0}}{\mu ^{4}}B^{a}B^{b}\left[ -x\, g_{ab}-\frac{\beta _{a}\beta _{b}}{\mu ^{2}}\right] 
\end{equation}

\subsection{Nonlinear approach to equilibrium in the homogeneous case}

We shall conclude this paper by studying the nonlinear evolution of the simple
model just described. By simplicity, we shall work in flat space time and assume
a homogeneous model, implying in particular that \( \beta ^{a} \) and \( B^{a} \)
are colinear, \( \beta ^{a}=\beta u^{a} \) and \( B^{a}=Bu^{a} \). We choose
a frame at rest with the fluid where \( u^{a}=\delta _{0}^{a} \) . 

The equations of motion are 
\begin{equation}
\label{un}
j^{a}_{\: ;a}=V^{\prime }(\phi )+\Delta 
\end{equation}

\begin{equation}
\label{deux}
T^{ab}_{+;a}=0
\end{equation}

\begin{equation}
\label{trois}
T^{ab}_{-;a}=2I^{b}
\end{equation}

\begin{equation}
\label{quatre}
\beta _{(c)}^{a}\phi _{,a}=-\xi 
\end{equation}

Let´s introduce the two energy-momentum tensor of both (perfect) fluids without
interaction 
\[
T_{c0}^{ab}=\frac{\xi ^{2}}{\mu ^{4}}\beta _{c}^{a}\beta _{c}^{b}+g^{ab}\left( \frac{1}{2}\frac{\xi ^{2}}{\mu _{c}^{2}}-V\left( \phi \right) \right) \]

and 
\[
T_{q0}^{ab}=g^{ab}p_{q}+u^{a}u^{b}\left( \rho _{q}+p_{q}\right) \]

The interaction between the two is given by (\( \beta =\frac{1}{2}(\beta _{c}+\beta _{q}) \)
and \( B=\beta _{c}-\beta _{q} \) ) 
\[
T^{ab}_{\Xi +}=\frac{\partial \Xi ^{a}}{\partial \beta _{b}}\]
 and 
\[
T^{ab}_{\Xi -}=2\frac{\partial \Xi ^{a}}{\partial B_{b}}\]

We use as model for our interaction term 
\[
\Xi =\frac{B^{2}}{\beta ^{4}}\left( F_{0}+G_{0}\right) \equiv \frac{B^{2}}{\beta ^{4}}\Gamma \]

Under the simplifying hypothesis stated above we have 
\[
T^{00}_{c0}=\frac{1}{2}\frac{\xi ^{2}}{\beta _{c}^{2}}+V\left( \phi \right) \]

and 
\[
\frac{\partial \Xi ^{0}}{\partial \beta _{0}}=20\frac{B^{2}}{\beta ^{6}}\Gamma \]

\[
\frac{\partial \Xi ^{0}}{\partial B_{0}}=-8\frac{B}{\beta ^{5}}\Gamma \]

It is convenient to introduce a new variable \( s \) defined as 
\[
B=\beta _{c}-\beta _{q}=s\beta _{q}\]
 implying 
\[
\beta =\frac{\beta _{c}+\beta _{q}}{2}=\left( \frac{s}{2}+1\right) \beta _{q}\]

and to express the equations in terms of the canonical momentum 

\[
\pi =\phi _{,t}=-\frac{\xi }{\beta _{c}}\]

The energy-momentum tensors thus read 
\[
T_{+}^{00}=\frac{\pi ^{2}}{2}+V(\phi )+\left[ \sigma +20\Gamma \frac{s^{2}}{\left( 1+s/2\right) ^{6}}\right] \frac{1}{\beta _{q}^{4}}\]

\[
T_{-}^{00}=\frac{\pi ^{2}}{2}+V\left( \phi \right) -\left[ \sigma +16\Gamma \frac{s}{\left( 1+s/2\right) ^{5}}\right] \frac{1}{\beta _{q}^{4}}\]

We are left with four O.D.E. 
\[
-\pi _{,t}=V^{\prime }(\phi )+\Delta \]

\[
\left( T^{00}_{+}\right) _{,t}=0\]

\[
\left( T^{00}_{-}\right) _{,t}=2I^{0}\]

\[
\phi _{,t}=\pi \]

where 
\[
\Delta =M^{\xi \xi }\xi -2B\left( 1-\kappa \right) M^{\xi 0}\]

and 
\[
I^{0}=-BM^{00}+2\kappa \xi M^{\xi 0}\]

The dimensions of the \( M^{AB} \) terms are 
\begin{eqnarray*}
\left[ M^{\xi \xi }\right]  & = & T^{2}\\
\left[ M^{00}\right]  & = & T^{6}\\
\left[ M^{\xi 0}\right]  & = & T^{4}\\
\left[ \xi \right]  & = & T
\end{eqnarray*}

This validates the following modelling for the \( M^{AB} \)
\begin{eqnarray*}
M^{\xi \xi } & = & -\frac{K_{0}}{\tau ^{2}}\\
M^{00} & = & -\frac{L_{0}}{\tau ^{6}}\\
M^{\xi 0} & = & \frac{M_{0}}{\tau ^{4}}
\end{eqnarray*}

where \( \tau  \) is some function of the dynamical variables with dimensions
\( T^{-1} \) and \( K_{0},L_{0}>0 \). Moreover, positive entropy production
imposes the following relationship: 
\[
K_{0}L_{0}\geq M_{0}^{2}\]
 We will also take the \( q \) -fluid as describing radiation so that \( \rho =\sigma T^{4} \)
with the Stefan - Boltzmann constant \( \sigma >0. \) The equation (\ref{un})
becomes 
\[
-\pi _{,t}=V^{\prime }(\phi )+\frac{K_{0}}{\tau ^{2}}\left( s+1\right) \beta _{q}\pi -2\left( 1-\kappa \right) s\beta _{q}\frac{M_{0}}{\tau ^{4}}\]

Observe that 
\begin{eqnarray*}
\left[ \frac{1}{2}\pi ^{2}+V\left( \phi \right) \right] _{,t} & = & \pi \left[ \pi _{,t}+V^{\prime }\left( \phi \right) \right] \\
 & = & \pi \left[ -\frac{K_{0}}{\tau ^{2}}\left( s+1\right) \beta _{q}\pi +2B\left( 1-\kappa \right) \frac{M_{0}}{\tau ^{4}}\right] 
\end{eqnarray*}

\[
I^{0}=L_{0}\frac{s}{\tau ^{6}}\beta _{q}-\kappa \left( s+1\right) \frac{\beta _{q}}{\tau ^{4}}\pi M_{0}\]

And define

\[
K=K_{0}\left( s+1\right) \frac{\beta _{q}}{\tau ^{2}}\pi -2\left( 1-\kappa \right) s\beta _{q}\frac{M_{0}}{\tau ^{4}}\]

\[
c_{1}=\sigma +16\Gamma \frac{s}{\left( 1+s/2\right) ^{5}}\]

\[
c_{2}=\sigma +20\Gamma \frac{s^{2}}{\left( 1+s/2\right) ^{6}}\]

Then 
\[
\pi _{,t}=-V^{\prime }\left( \phi \right) -K\]

\[
-K\pi -c_{2}\frac{4}{\beta _{q}^{5}}\beta _{q,t}+\frac{40\Gamma }{\beta _{q}^{4}}\frac{s(1-s)}{\left( 1+s/2\right) ^{7}}s_{,t}=0\]

and 
\[
-K\pi +c_{1}\frac{4}{\beta _{q}^{5}}\beta _{q,t}-\frac{16\Gamma }{\beta _{q}^{4}}\frac{(1-2s)}{\left( 1+s/2\right) ^{6}}s_{,t}=2I^{0}\]

The final system of equations is 
\[
s_{,t}=\frac{\beta _{q}^{4}\left( 1+s/2\right) ^{6}}{8\Gamma c_{3}}\left\{ 2I^{0}c_{2}+K\pi \left( c_{1}+c_{2}\right) \right\} \]

\[
\beta _{q,t}=\frac{\beta _{q}^{5}}{4c_{3}}\left\{ 10\frac{s\left( 1-s\right) }{1+s/2}I^{0}+K\pi h\left( s\right) \right\} \]

\[
\pi _{,t}=-V^{\prime }(\phi )-K\]

\[
\phi _{,t}=\pi \]

where 

\[
c_{3}=5\frac{s\left( 1-s\right) }{\left( 1+s/2\right) }c_{1}-2\left( 1-2s\right) c_{2}\]

\[
h\left( s\right) =2\left( 1-2s\right) +\frac{5(1-s)s}{1+s/2}\]

\[
V\left( \phi \right) =\frac{1}{2}m^{2}\phi ^{2}+\lambda \phi ^{4}\]

Finally 
\[
T^{00}_{+}=\frac{1}{2}\pi ^{2}+V\left( \phi \right) +\frac{c_{2}}{\beta _{q}^{4}}\]

is both positive definite and conserved. 

These equations generally describe the approach to equilibrium. This is most
clearly seen in the limit where \( s \) remains small. In this limit, we have
\( c_{1}\sim c_{2}\sim \sigma  \), \( c_{3}\sim -2\sigma <0 \), \( h\left( s\right) \sim 2 \)
and \( K\sim K_{0}\beta _{q}\pi /\tau ^{2}. \) In particular, the equation
for the field \( \phi  \) describes damped oscillations, but the damping ''constant''
is truly a dynamical variable, thus opening a mechanism to include memory effects
in the dynamics and interesting behavior; note by exemple that \( K \) can
change sign if the coupling term \( M_{0}\neq 0. \)

More generally, by exploring parameter space we find a variety of behaviors.
In particular, the approach to equilibrium may be either over or underdamped;
this second case seems to be relevant to the description of preheating episodes.

\section{Final Remarks}

The main conclusion of this work is that the conditions of causality, stability
and a proper thermodynamic behavior put concrete limits on possible phenomenological
models of the reheating period. We have shown concrete examples of divergence
type theories which satisfy these requirements. Unlike earlier work, we have
described the inflaton as a field, rather than disregarding its coherence by
describing it simply as another fluid. This has required an extension both of
the usual Klein - Gordon and DTT frameworks. We have also shown how this field
can be consistently coupled to a fluid. 

The equations of motion we have derived in this last Section show also that
it is possible to explore the bulk of possible dynamical behaviors already with
models with a minimal set of undetermined parameters (in our case, these were
\( \Gamma  \), \( K_{0} \), \( L_{0} \), \( M_{0} \) and the functional
form of \( \tau  \)). These parameters may be estimated by fitting the predictions
of the model to microscopic calculations in controlled limiting cases, much
in the same way as viscosity coefficients in field theory are computed by analyzing
the damping of extra long wavelength fluctuations \cite{Kadanoff,J1,J2,JeonYaffe,Carrington99,CalzettaHu}.
We may then obtain reliable phenomenological models to use as a tool to explore
the full nonlinear physics of reheating, with an enormous gain in simplicity
as compared to a full attack from a first principles perspective.

\section{Appendix: Divergence Type Theories}

Following Geroch \cite{Geroch}, divergence type theories are usually described
in terms of some tensorial quantities that obey conservation equations 
\begin{eqnarray*}
T_{\; ;b}^{ab} & = & 0\\
N_{\; ;a}^{a} & = & 0\\
A_{\quad ;a}^{abc} & = & I^{bc}
\end{eqnarray*}

This is a simple and slight generalization of relativistic fluid theories proposed
initially by Liu, Muller and Ruggeri \cite{Liu}. In this setting, \( T^{ab} \)
is the energy-momentum tensor and \( N^{a} \) is the particle current. Their
corresponding equation simply expresses conservation of energy, momentum and
mass. The third equation will describe the dissipative part. The energy-momentum
tensor is symmetric and \( A^{abc}=A^{acb}; \) \( A_{\quad b}^{ab}=0 \) and
\( I_{\, \, ;a}^{a}=0. \) The entropy current is enlarged to read 
\[
S^{a}=\chi ^{a}-\xi _{b}T^{ab}-\xi N^{a}-\xi _{bc}A^{abc}\]

The \( \xi ,\xi ^{a},\xi ^{ab} \) are the dynamical degrees of freedom. The
following relations hold \cite{Geroch} 
\[
N^{a}=\frac{\partial \chi ^{a}}{\partial \xi }\]

\[
T^{ab}=\frac{\partial \chi ^{a}}{\partial \xi _{b}}\]

\[
A^{abc}=\frac{\partial \chi ^{a}}{\partial \xi _{bc}}\]

Symmetry of the energy-momentum tensor implies that 
\[
\chi ^{a}=\frac{\partial \chi }{\partial \xi _{a}}\]

That is all the fundamental tensors of the theory can be obtained from the generating
functional \( \chi . \) The entropy production is given by 
\[
S_{\, \, ;a}^{a}=-I^{bc}\xi _{bc}\]

Positive entropy production is ensured by demanding that \( I^{bc}=M^{(bc)(de)}\xi _{de} \),
where \( M \) is negative definite. 

Ideal fluids are an important if somewhat trivial example. To obtain ideal hydrodynamics
within the DTT framework, consider a generating functional \( \chi _{p}=\chi _{p}\left( \xi ,\mu \right)  \)
where \( \mu \equiv \sqrt{-\xi _{a}\xi ^{a}}. \) It is a simple matter to obtain
\[
\chi _{p}^{a}=-\frac{\xi ^{a}}{\mu }\frac{\partial \chi _{p}}{\partial \mu }\]

\[
T_{p}^{ab}=-\frac{g^{ab}}{\mu }\frac{\partial \chi _{p}}{\partial \mu }+\frac{\xi ^{a}\xi ^{b}}{\mu ^{2}}\left[ -\frac{1}{\mu }\frac{\partial \chi _{p}}{\partial \mu }+\frac{\partial ^{2}\chi _{p}}{\partial \mu ^{2}}\right] \]

A simple comparison with the perfect fluid form of the energy-momentum tensor
\( T^{ab}=g^{ab}p+u^{a}u^{b}\left[ p+\rho \right]  \) implies the following
identification 
\begin{equation}
\label{pressure}
p=-\frac{1}{\mu }\frac{\partial \chi _{p}}{\partial \mu }
\end{equation}

\[
\rho =\frac{\partial ^{2}\chi _{p}}{\partial \mu ^{2}}\]

Note that the conserved current can be quite generally written as 
\begin{equation}
\label{current}
N^{a}=\frac{\partial }{\partial \xi }\left( -\frac{\xi ^{a}}{\mu }\frac{\partial \chi }{\partial \mu }\right) =\xi ^{a}\frac{\partial p}{\partial \xi }
\end{equation}

A less trivial but important example both historically and conceptually is the
Eckart theory which can be obtained from \cite{Geroch} 
\[
\chi _{E}=\chi _{p}+\frac{1}{2}\zeta _{ab}u^{a}u^{b}\]

Performing a Legendre transform to the new variables \( \xi \, \xi ^{a},\xi ^{ab} \)
one obtains a system of first order differential equations of the form 
\[
\frac{\partial ^{2}\chi ^{a}}{\partial \xi _{A}\partial \xi _{B}}\xi _{B;a}=I^{A}\]

where \( \xi _{A} \) stand for the entire collection of variables \( \left( \xi ,\xi _{a}\, ,\xi _{ab}\right)  \)
and similarly \( I^{A}\equiv \left( 0,0,I^{ab}\right)  \) represent the dissipative
source; the index \( A \) thus covers \( 14 \) dimensions in our example.
This first order system of differential equations is symmetric since 
\[
\frac{\partial ^{2}\chi ^{a}}{\partial \xi _{A}\partial \xi _{B}}=\frac{\partial ^{2}\chi ^{a}}{\partial \xi _{B}\partial \xi _{A}}\]

Note that we have a system of the form 
\[
A^{i}v_{,i}+Bv=0\]

where \( i \) is a space-time index, the \( A^{i} \) and \( B \) are \( k\times k \)
matrices and \( v \) is a \( k \) -vector. Now this (first order) system is
hyperbolic if all its eigenvalues are real; each of these eigenvalues represent
the velocity of propagation of some small disturbance in space. These in turn
propagate along hypersurfaces called characteristics whose existence is insured
by the existence of \( k \) real eigenvalues \cite{Courant,Stewart}. If the
matrices \( A^{i} \) and \( B \) are symmetric then it suffices that some
combination \( A^{i}v_{i} \) be definite (negative-definite given our choice
of the signature for the metric) to insure that all the eigenvalues are real
(but some could be degenerate). An usual case happens when this combination
reduces to \( A^{0} \) , the vector \( v \) being the time-like vector \( \left( 1,\overrightarrow{0}\right) . \)
In a relativistic theory one would expect hyperbolicity to be invariant under
(proper) Lorentz transformations; in this case we say the system is causal.
In our context, one would thus say that the system is hyperbolic if 
\[
\frac{\partial ^{2}\chi ^{a}}{\partial \xi _{A}\partial \xi _{B}}v_{a}\]
 is negative-definite for some temporal vector \( v_{a} \) and the theory will
be causal if this stay true for \emph{any} temporal vector \( v_{a}. \)

\section{Acknowledgments}

This work has been partially supported by Universidad de Buenos Aires, CONICET,
Fundaci\'{o}n Antorchas and the ANPCYT through project PICT99 03-05229. A preliminary
version of this work was presented at the ReBReG 2000 (La Plata, Argentina,
December 2000); we thank the organizers H. Vucetich and S. Landau for their
invitation, as well as O. Reula, O. Ortiz and J. Geron for useful discussions.

\end{document}